\documentclass[12pt]{iopart}
\usepackage{iopams}
\usepackage{graphicx}

\def\d{\delta}
\def\m{\mu}
\def\n{\nu}
\newcommand{\be}{\begin{equation}}
\newcommand{\ee}{\end{equation}}
\newcommand{\nbe}{\begin{equation*}}
\newcommand{\nee}{\end{equation*}}

\begin{document}
\title[New one-parametric extension of the Starobinsky inflationary model]{\textbf{New one-parametric extension of the Starobinsky inflationary model}}

\author{Ekaterina~O.~Pozdeeva}
\address{Skobeltsyn Institute of Nuclear Physics, Lomonosov Moscow State University, Leninskie Gory~1, Moscow, 119991,  Russia}
\ead{pozdeeva@www-hep.sinp.msu.ru}
\author{Sergey~Yu.~Vernov}
\address{Skobeltsyn Institute of Nuclear Physics, Lomonosov Moscow State University, Leninskie Gory~1, Moscow, 119991,  Russia}
\ead{svernov@theory.sinp.msu.ru}

\date{ \ }
\begin{abstract}
We propose a one-parametric extension of the Starobinsky $R+R^2$ model by adding the $\left(R+m^2\beta^2\right)^{3/2}$ term. The parameter $m$ is the inflaton mass, which is determined in the same way as in the Starobinsky model, and $\beta$ is a dimensionless constant. Using the Einstein frame and the scalar field potential, we get the inflationary parameters of the model proposed.
The value of the tensor-to-scalar ratio $r$ can be significantly larger than in the Starobinsky model. The considered inflationary model is in a good agreement with the current observational data. The corresponding scalar field potential is a polynomial of the exponential function.
\end{abstract}

\vspace{2pc}

\vspace{2pc}




\section{Introduction}
\label{sec:intro}
The simplest and the most popular extension to the General relativity (GR) is the $F(R)$ gravity~\cite{Sotiriou:2008rp,DeFelice:2010aj,Capozziello:2011et,Nojiri:2017ncd,Bajardi:2022ocw} with  a nonlinear function $F(R)$ in the action.  In the context of $F(R)$ gravity, inflationary scenarios are being actively studied~\cite{Starobinsky:1980te,Mukhanov:1981xt,Starobinsky:1983zz,Vilenkin:1985md,Mijic:1986iv,Maeda:1987xf,Barrow:1988xh,Berkin:1990nu,Saidov:2010wx,Huang:2013hsb,Sebastiani:2013eqa,Motohashi:2014tra,Broy:2014xwa,Motohashi:2014ppa,Bamba:2015uma,Odintsov:2016imq,Miranda:2017juz,
Motohashi:2017vdc,Paliathanasis:2017apr,Mishra:2019ymr,Cheong:2020rao,Rodrigues-da-Silva:2021jab,Ivanov:2021chn,Odintsov:2022rok,Odintsov:2022bpg,Modak:2022gol}.

Historically, the first $F(R)$ gravity inflationary model is the Starobinsky $R+R^2$ model~\cite{Starobinsky:1980te}, described by the following action:
\be
\label{starm}
S_{\rm Star.} = \frac{M^2_{Pl}}{2}  \int d^4x\sqrt{-g} \left(R
+ \frac{1}{6m^2} R^2\right)~,
\ee
where the reduced Planck mass $M_{Pl}$ and the inflaton mass $m$ are introduced. The Ricci scalar $R$ is defined by the metric
$g_{\m\n}$ with the spacetime signature $(-,+,+,+)$.

The Starobinsky model is known as an excellent model of the large-field slow-roll cosmological inflation~\cite{Starobinsky:1980te,Mukhanov:1981xt,Starobinsky:1983zz,Vilenkin:1985md,Mijic:1986iv,Maeda:1987xf}. A good agreement of the predictions of this model to the Planck measurements of the Cosmic Microwave Background (CMB) radiation has been noted in Ref.~\cite{Planck:2018jri}. Moreover, the recent combined analysis of the observation data confirms this statement~\cite{BICEP:2021xfz,Tristram:2021tvh}. The inflaton mass $m$ is fixed by CMB measurements of the amplitude of scalar perturbations $A_s$. Note that the values of the scalar spectral index $n_s$ and the tensor-to-scalar ratio $r$ do not depend on $m$. This model is the simplest inflationary model with the maximum predictive power. As known, the value of tensor-to-scalar ratio $r$ is  restricted only from above~\cite{BICEP:2021xfz,Tristram:2021tvh}. If the observed by future experiments value of the tensor-to-scalar ratio $r$ differs from its value in the Starobinsky model, then some corrections of this model will be required.

There are two main ways to generalize the Starobinsky model without adding a scalar fields or other matter. One can either add string theory inspired terms~\cite{Ketov:2010eg,Ketov:2010qz,Ketov:2012se,Koshelev:2016xqb,Koshelev:2020foq,Ketov:2022lhx,Koshelev:2022olc,Rodrigues-da-Silva:2022qiq,Ketov:2022zhp,Bezerra-Sobrinho:2022dkv} or construct new $F(R)$ gravity inflationary models, connected to the Starobinsky model~\cite{Barrow:1988xh,Berkin:1990nu,Saidov:2010wx,Huang:2013hsb,Sebastiani:2013eqa,Broy:2014xwa,Motohashi:2014tra,Bamba:2015uma,
Odintsov:2016imq,Miranda:2017juz,Cheong:2020rao,Rodrigues-da-Silva:2021jab,Ivanov:2021chn,Odintsov:2022rok,Modak:2022gol}. Also, it is possible to combine these two ways, obtaining $F(R)$ models related to fundamental theories of gravity.

The $\mathcal{F}(\mathcal{R})$ supergravity models with a holomorphic function $\mathcal{F}$ of the covariantly-chiral scalar curvature superfield $\mathcal{R}$ have been proposed and studied in Refs.~\cite{Ketov:2010eg,Ketov:2010qz,Ketov:2012se}. These models can be considered as the $N = 1$ locally supersymmetric extension of the $F(R)$ gravity.
The supergravity model with a quadratic $\mathcal{F}(\mathcal{R})$ function, where the linear term is supposed to represent the standard (pure) $N = 1$ Poincar\'e supergravity and the quadratic term is considered as a 'quantum correction', gives the following $F(R)$ gravity action~\cite{Ketov:2010eg}:
\begin{equation}
\label{ActionSFR}
    S_{\mathcal{F}}=\frac{85M_{Pl}^2}{198}R -\frac{14M^2_{Pl}}{99} (R + R_0)\left(1 \pm \sqrt{1 + \frac{R}{R_0}}\right)\,,
\end{equation}
where $R_0$ is a constant.

In Refs.~\cite{Ketov:2010qz,Ketov:2012se}, it has been shown that the $R^{2}$ term can be obtained from the supergravity model with a cubic $\mathcal{F}(\mathcal{R})$ function. The obtained $F(R)$ gravity model is not the Starobinsky one, because it includes also the $(R + R_0)^{3/2}$ term. In the limit $|R|\ll R_0$, one gets the model with the $R^{3/2}$ term that is suitable for inflation~\cite{Ketov:2012se}. In Ref.~\cite{Ivanov:2021chn}, it has been shown that  adding of the $R^{3/2}$ term allows to construct a viable inflationary model with the tensor-to-scalar ratio $r$ four times larger
than in the Starobinsky model~\cite{Starobinsky:1980te,Starobinsky:1983zz}.

Note that the $R^{3/2}$-term also appears in an approximate description of the Higgs field with a small cubic term in its potential and a large non-minimal coupling to~$R$~\cite{Martins:2020oxv}.
Cosmological models with $R^{\gamma}$ and $(R+R_0)^{\gamma}$ terms, where $\gamma$ is not integer,  are actively investigated~\cite{Bajardi:2022ocw,Sebastiani:2013eqa,Motohashi:2014tra,Odintsov:2022bpg,Capozziello:1999xs,Carloni:2004kp,Capozziello:2008ch,Martin-Moruno:2008qpc,Sarkar_2013,Modak:2014yza,Paliathanasis:2016tch,Vernov:2019ubo,Leon:2022dwd}. Models with the $R^{3/2}$ term have been investigated in context of integrability and symmetries~\cite{Capozziello:1999xs,Capozziello:2008ch,Sarkar_2013,Paliathanasis:2016tch}, in particular, it has been shown in Refs.~\cite{Capozziello:1999xs,Capozziello:2008ch,Sarkar_2013} that the $R^{3/2}$ term appears naturally from the Noether symmetry analysis.

To avoid the graviton as a ghost and the scalaron (inflaton) as a tachyon in a $F(R)$ gravity model one should put the following conditions~\cite{Starobinsky:2007hu,Appleby:2009uf}:
\be \label{restr2}
F_{,R}(R) >0 \quad {\rm and} \quad F_{,RR}(R)>0\,,
\ee
where $F_{,R}$ and $F_{,RR}$ are the first and the second derivatives of $F(R)$ with respect to $R$, correspondingly. So, the Starobinsky model~\cite{Starobinsky:1980te} is well-defined and has no ghost for all $R>-3m^2$.

At the same time, any model with the $R^{3/2}$ term is ill-defined at $R<0$. In the Starobinsky model, the scalar curvature $R$  monotonically decreases during inflation and oscillates near zero with a decreasing amplitude after inflation~\cite{Arbuzova:2011fu,Arbuzova:2021etq}. In model with the $R^{3/2}$ term, we cannot reproduce such oscillations of $R$.
By this reason, it is interesting to consider model with the $(R + R_0)^{3/2}$ term that is well-defined for some negative values of $R$.

In  this paper, we consider a new inflationary model with the $(R+R_0)^{3/2}$-term with a positive constant $R_0=\beta^2 m^2$. The proposed function
\begin{equation}
\label{FR32alpha}
    F(R)=\frac{M^2_{Pl}}{2}\left[ \left(1-\frac{3}{2}\beta\delta\right)R + \frac{R^2}{6m^2}  + \frac{\d}{m}\left(R+\beta^2 m^2\right)^{3/2}-m^2\beta^3\d\right],
\end{equation}
includes two dimensionless parameters $\d$ and $\beta$. At small values of $R/m^2$, the model should reproduce the General Relativity action. By this reason, we add the cosmological constant term and modify the linear term.
We demonstrate that the proposed model is well-defined and its first and second derivatives are positive for all $R>R_{min}$, where $R_{min}<0$. To construct a new one-parametric generalization of the Starobinsky model we connect the parameters $\d$ and $\beta$ (see Section~\ref{Model}). The considered model is a straightforward generalization of action (\ref{ActionSFR}) and includes the $R^2$ term.

The proposed method is different from the way using in Ref.~\cite{Ivanov:2021chn}. We start from the model that includes two new dimensionless parameters, after this we find the connection between these parameters to simplify the form of the corresponding potential in the Einstein frame. This way allows us to get such a one-parameter extension of the Starobinsky model  (\ref{starm}) that the corresponding potential is a polynomial of the exponential function of a canonical scalar field for any values of the model parameters.

\section{$F(R)$ models and the corresponding scalar potentials}

The generic $F(R)$ gravity model has the following action
\begin{equation}
\label{f_R_action}
    S_F=\int d^4 x \sqrt{-g}F(R),
\end{equation}
with a differentiable function $F$.

Varying action (\ref{f_R_action}), one gets the following equations:
\begin{equation}
F_{,R}{R}^\mu_\nu-\frac{1}{2} F\delta^\mu_\nu- \left(g^{\mu\rho}\nabla_\rho\nabla_\nu-\delta^\mu_{\nu}
\Box\right)F_{,R}=0,
\label{equ_f_R}
\end{equation}
where the covariant d'Alembertian $\Box=g^{\alpha\beta}\nabla_\alpha\nabla_\beta$.

The $F(R)$ gravity action (\ref{f_R_action}) can be rewritten as follow:
\begin{equation}
\label{SJ}
    S=\int d^4 x \sqrt{-g} \left[ F_{,\sigma} (R-\sigma)+ F\right],
\end{equation}
where a new scalar field $\sigma$ has been introduced, and $F_{,\sigma}(\sigma)=\frac{dF(\sigma)}{d\sigma}$~.
If $F_{,\sigma\sigma}(\sigma)\neq 0$, then varying (\ref{SJ}) over $\sigma$ and eliminating $\sigma$ via the obtained equation $R=\sigma$, one yields back the action~(\ref{f_R_action}).

After the Weyl transformation of the metric
\begin{equation*}
g^E_{\mu\nu}=\frac{2F_{,\sigma}(\sigma)}{M^2_{Pl}}g_{\mu\nu},
\end{equation*}
one gets the following action in the Einstein frame~\cite{Maeda:1988ab}:
\begin{equation}
\label{SE}
S_{E} =\int d^4x\sqrt{-g^E}\left[\frac{M^2_{Pl}}{2}R_E-\frac{h(\sigma)}{2}{{g^E}^{\mu\nu}}\partial_\mu{\sigma}\partial_\nu{\sigma}-V_E({\sigma})\right],
\end{equation}
where $R_E$ is defined by the metric $g^E_{\m\n}$,
\begin{equation}
\label{Ve}
h(\sigma)=\frac{3M^2_{Pl}}{2F_{,\sigma}^2}F_{,\sigma\sigma}^2 \quad {\rm and}\quad
V_E(\sigma)= \frac{M^4_{Pl}}{4F_{,\sigma}^2}\left(F_{,\sigma} \sigma-F\right).
\end{equation}
Note that the considering Weyl transformation is well-defined for $F_{,\sigma}>0$ only.

One can get the action $S_{E}$ in the following form:
\begin{equation}\label{ActionSe}
    S_E =\int  d^4x \sqrt{-g^E}\left[\frac{ M^2_{Pl}}{2}R_E-\frac{1}{2}{g^E}^{\m\n}\partial_\mu\phi\partial_\nu\phi-V(\phi)\right],
\end{equation}
where the canonical scalar field
\begin{equation}
\label{phidF}
\phi=\sqrt{\frac{3}{2}}M_{Pl}\ln\left[\frac{2}{M^2_{Pl}}F_{,\sigma}(\sigma)\right]\,.
\end{equation}
It has been shown in~\cite{Ivanov:2021chn}, that the variable
\be \label{ydef}
y \equiv \exp\left( -\sqrt{\frac{2}{3}}\frac{\phi}{M_{Pl}}\right) =\frac{M^2_{Pl}}{2F_{,\sigma}(\sigma)}
\ee
is useful for considering of generalization of the  Starobinsky inflationary model, because it is small during inflation.

In the case of the Starobinsky model, the scalar potential $V(y)$ has the following form:
\be \label{starpd}
V_{\rm Star.}(y)= V_0\left(1-y\right)^2~,\quad {\rm where} \quad V_0=\frac{3}{4}m^2M_{Pl}^2~~.
\ee

In the slow-roll approximation, the evolution of the scalar field is defined by the following equation (see~\cite{Ivanov:2021chn} for detail):
\begin{equation}
\label{equyslr}
    y'=\frac{2y^2V_{,y}}{3V}~.
\end{equation}
where the prime denotes the derivative with respect to e-folding number
\begin{equation*}
N_e=\ln \left( \frac{a_{\mathrm{end}}}{a} \right)~.
\end{equation*}
We fix $a_{\rm end}$ as the value of $a$ at the end of inflation, so the exit from inflation appears at $N_e=0$.

The slow-roll parameters are defined as follows:
\begin{equation}\label{slrparam}
\epsilon=\frac{y^2}{3}\left(\frac{V_{,y}}{V}\right)^2\,,\qquad
\eta=\frac{2y}{3V}\left(V_{,y}+yV_{,yy}\right)~.
\end{equation}

The main cosmological parameters of inflation are given by the scalar spectral index $n_s$ and the tensor-to-scalar ratio $r$,
whose values are constrained by the combined Planck, WMAP and BICEP/Keck observations~\cite{Planck:2018jri,BICEP:2021xfz,Tristram:2021tvh}:
\be
\label{PlanckCMB}
n_s=0.9649\pm 0.0042  \quad ({\rm 68\% CL}) \qquad {\rm and} \qquad  r < 0.036 \quad ({\rm 95\% CL})~.
\ee

In the slow-roll approximation, these parameters are connected with the slow-roll parameters as follows~\cite{Liddle:1994dx}:
\begin{equation}
\label{nsr}
    n_s=1-6\epsilon+2\eta,\qquad r=16\epsilon~.
\end{equation}
The amplitude of scalar perturbations is given by
\be
\label{As}
A_s=\frac{2V}{3\pi^2M_{Pl}^4 r}\,.
\ee
Its value observed by the Planck telescope~\cite{Planck:2018jri} is $A_s=2.1\times 10^{-9}$.

\section{The generic inflationary model with the $(R+R_0)^{3/2}$ term}
\label{Model}

We consider the model with the function $F(R)$ given by (\ref{FR32alpha}). At $\delta \geqslant 0$,  the first derivative
\begin{equation}\label{DFR}
F_{,R}(R)=\frac{M^2_{Pl}}{4}\left(2-3\delta\beta\right)+\frac{M^2_{Pl}}{6m^2}R+\frac{3M^2_{Pl}\delta}{4m}\sqrt{\beta^2 m^2+R} >\frac{M^2_{Pl}}{2}
\end{equation}
for any $R> 0$ as well as in the Starobinsky model. Note that $F_{,R}(0)=M^2_{Pl}/{2}$ for all $\beta\geqslant 0$.

The second derivative
\begin{equation}
\label{D2F}
F_{,RR}(R)=\frac{M^2_{Pl}}{6m^2}+\frac{3M^2_{Pl}\delta}{8m\sqrt{\beta^2m^2+R}}>0
\end{equation}
for any $R>-\beta^2 m^2$ and $\d\geqslant 0$.

 So, at $\beta\neq 0$ and $\d\geqslant 0$, the model (\ref{FR32alpha}) is well-defined for all $R>R_{min}$, where $-\beta^2m^2\leqslant R_{min}<0$.

In the case of $\beta\neq 0$, the function $F(R)$ has a correct GR limit at $R\ll m^2$:
\begin{equation}
\label{FRsmall}
F=\frac{M^2_{Pl}}{2}R\left[1+\left(1+\frac{9\delta}{4\beta}\right)\frac{R}{6m^2}+{\cal{O}}\left(\frac{R^2}{m^4}\right)\right],
\end{equation}
more exactly,
\begin{equation}
\label{Fserias}
 F=\frac{M^2_{Pl}}{2}R\left[1+\left[\frac{1}{6}+\frac{3\delta}{8\beta}\right] \tilde{\sigma}-\frac{\delta}{16\beta^3}\tilde{\sigma}^2+\frac{3\delta}{128\beta^5}\tilde{\sigma}^3+{\cal{O}}\left({\tilde{\sigma}}^4\right)\right]\,,
\end{equation}
where $\tilde{\sigma}=R/m^2$.

To get inflationary parameters we construct the corresponding potential~(\ref{Ve}):
\begin{equation}
\label{V32alpha}
    V_E(\tilde{\sigma})=\frac{4V_0\left(6\beta^3\delta+3\delta\sqrt{\beta^2+\tilde{\sigma}}\left(\tilde{\sigma}-2\beta^2\right)+\tilde{\sigma}^2\right)}{\left(9\delta\beta-9\delta\sqrt{\beta^2+\tilde{\sigma}}-2\tilde{\sigma}-6\right)^2}\,,
\end{equation}

At $\delta=0$, we get the potential for the Starobinsky inflationary model:
\begin{equation}
\label{VStarsigma}
    V_{\mathrm{Star.}}=\frac{V_0\tilde{\sigma}^2}{\left(\tilde{\sigma}+3\right)^2}\,.
\end{equation}

To obtain $V(y)$ we solve Eq.~(\ref{ydef}):
\begin{equation}\label{ysigmaequ}
   y+\frac{6}{9\delta\beta-9\delta\sqrt{\beta^2+\tilde{\sigma}}-2\tilde{\sigma}-6}=0,
\end{equation}
and get
\begin{equation}
\label{sigmay}
\tilde{\sigma}_\pm=\frac{3(1-y)}{y}+\frac{9\delta}{8y}\left(4\beta y+9\delta y\pm s\right),
\end{equation}
where $s=\sqrt{\left[72\delta\beta +81\delta^2+16\beta^2 -48\right]y^2+48y}$.
To get $\sigma=0$ at $y=1$ for all nonnegative values of $\delta$ and $\beta$ we choose solution $\tilde{\sigma}_-$.

\section{A new one-parametric generalization of the Starobinsky inflationary model}
In the general case, the corresponding scalar field potential $V(y)$ is rather complicate.
To get $V(y)$ in a simple form, we connect the values of parameters $\beta$ and $\delta$ as follows:
\begin{equation}
\label{deltabeta1}
    \beta=\sqrt{3}-\frac{9}{4}\delta.
\end{equation}
In this case, $s=4\sqrt{3y}$ and
\begin{equation}
\label{sigmay1}
    \tilde{\sigma}_-(y)={}-\frac{3}{2y}\left(3\sqrt{3}\,\delta\sqrt{y}-3\sqrt{3}\,\delta\, y+2y-2\right)\,.
\end{equation}
The corresponding potential has the following form:
\begin{equation}
\label{Vy1}
V(y)=V_0\left[\left(y-1\right)^2-\sqrt{3}\delta\sqrt{y}\left(\sqrt{y}+2\right)\left(\sqrt{y}-1\right)^2\right],
\end{equation}
that is equivalent to
\begin{equation}\label{Vphi1}
    V(\phi)=V_{\mathrm{Star.}}+\sqrt{3}V_0\delta\mathrm{e}^{-{\phi}/(\sqrt{6}M_{Pl})}\left[\mathrm{e}^{-{\phi}/(\sqrt{6}M_{Pl})}+2\right]
    \left[\mathrm{e}^{-{\phi}/(\sqrt{6}M_{Pl})}-1\right]^2\!.
\end{equation}
So, the potential $V(\phi)$ is a polynomial of the exponential function.
Using
\begin{equation}\label{DVy1}
    V_{,y}(y)=\frac{\sqrt{3}V_0}{3\sqrt{y}}\left(2\,\sqrt{3}y-6\delta\,y+2
\sqrt{3}\sqrt{y}-6\delta\,\sqrt{y}+3\delta \right)\left(\sqrt{y}-1\right)\,,
\end{equation}
we get that the potential $V(y)$ always has an extremum at
$y=1$.  This point corresponds to $V=0$ and the post-inflationary evolution of the Universe. Other extreme correspond to
\begin{equation}
\label{y}
    \sqrt{y_{m_{\pm}}}={}-\frac{1}{2}\pm\frac{\sqrt{3-12\sqrt{3}\delta+27\delta^2}}{2\left(3\delta-\sqrt{3}\right)}.
\end{equation}

\begin{figure}[htb]
\centering
 \includegraphics[width=0.47\textwidth]{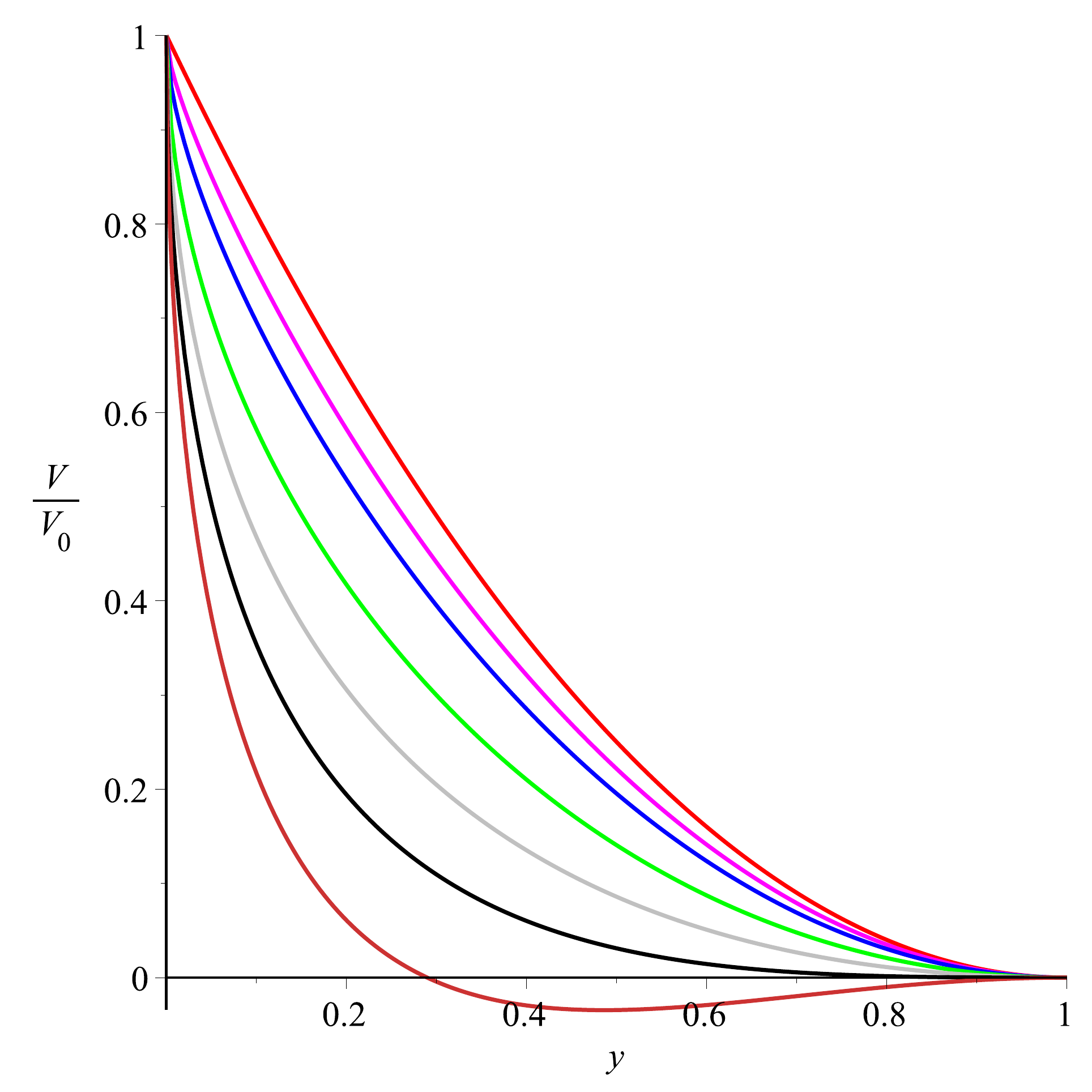} \ \ \
  \includegraphics[width=0.47\textwidth]{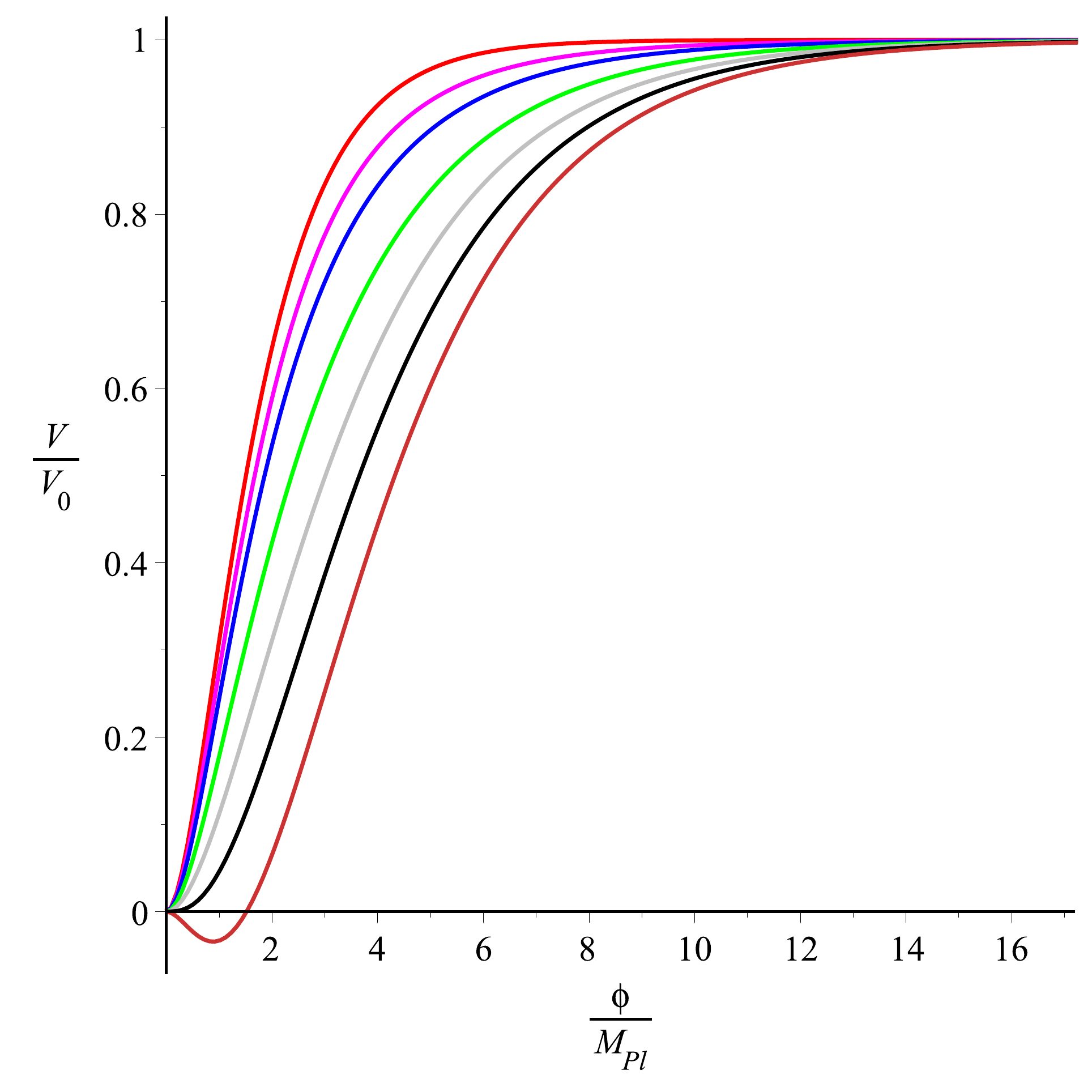}
\caption{The potentials $V(y)$ (left) and ${V}(\phi)$ (right)
in the case of $\beta=\sqrt{3}-\frac{9}{4}\delta$ for $\d=0$ (red), $\d=0.1$ (magenta), $\d=\sqrt{3}/9$ (blue), $\d=2\sqrt{3}/9$ (green), $\d=\sqrt{3}/3$ (grey), $\d=4\sqrt{3}/9$ (black), and $\d=1$ (orange).}
\label{F32oneparameterVphi}
\end{figure}

It is easy to see that $\sqrt{y_{m_{\pm}}}<0$ for $0<\delta\leqslant\sqrt{3}/9$ and $\sqrt{y_{m_{\pm}}}$ are not real for $\sqrt{3}/9<\delta< 1/\sqrt{3}$. At $1/\sqrt{3}\leqslant\delta$, the function $\sqrt{y_{m_{+}}}>0$ and the function $\sqrt{y_{m_{-}}}<0$. We get $0<y_{m_{+}}<1$ at $\d>4\sqrt{3}/9$ and $y_{m_{+}}=1$ at $\d=4\sqrt{3}/9$.
So, the potential $V(y)$ has no extremum in the interval $0<y<1$ if $0\leqslant\d\leqslant4\sqrt{3}/9$ that is equivalent to $\sqrt{3}\geqslant \beta\geqslant 0$. The potentials $V(y)$  and ${V}(\phi)$ for different values of $\delta$ are presented in Fig.~\ref{F32oneparameterVphi}. One can see that the potentials have minima in the case of $\delta=1$.

Using Eq.~(\ref{slrparam}), we get the slow-roll parameters as functions of $y$:
\begin{equation}
\label{eps32dg}
   \epsilon=\frac{y\left(\sqrt{y}-1\right)^2\left(2\sqrt{3}\,\delta\,y-2\,y+2\sqrt{3}\,\delta\,\sqrt{y}-2\,\sqrt{y}-\sqrt{3}\,\delta\right)^2}{3\left(\sqrt{3}\,\delta\left({y}^{2}-3y
   +2\sqrt{y}\right)-\left(y-1\right)^2 \right)^2}
\end{equation}
and
\begin{equation}
\label{eta32dg}
    \eta=\frac{\sqrt{y}\left(8\,\sqrt {3}\delta {y}^{3/2}-8\,{y}^{3/2}-6\,\sqrt {3}
\delta\,\sqrt {y}+4\,\sqrt {y}+\sqrt {3}\delta \right) }{3\left(\sqrt{3}\,\delta \left({y}^{2}-3y+2\sqrt{y}\right)-\left(y-1\right)^2\right)}.
\end{equation}

Therefore, according to Eq.~(\ref{nsr}), the inflationary parameters as functions of $y$ are
\begin{equation}
\label{nseta32dg}
\begin{array}{ccl}
    n_s(y)\!&=&\displaystyle  1-\frac{2}{3\left(\sqrt{3}\,\delta\left({y}^{2}-3y
   +2\sqrt{y}\right)-\left(y-1\right)^2 \right)^2}\\
&\displaystyle \times&\left(5\,\sqrt {3}\,\delta\,{y}^{5/2} - 15\,{\delta}^{2}{y}^{5/2}-8
\sqrt{3}\,\delta\,{y}^{4}+12\,{\delta}^{2}{y}^{4}+2\sqrt{3}\,\delta\,{y}^{3/2}
\right.\\
&{}+&\displaystyle \left.10\sqrt{3}\,\delta\,{y}^{3}-9{\delta}^{2}{y}^{3/2}-18{\delta}^{2}{y}^{3}
-4\sqrt {3}\,\delta\, {y}^{2}+27{\delta}^{2}{y}^{2}\right.\\
&+&\displaystyle \left.\sqrt{3}\,\delta\,\sqrt {y}-6\sqrt {3}\,\delta\,y+3{\delta}^{2}\,y+4\,{y}^
{4}-4
\,{y}^{3}-4\,{y}^{2}+4\,y\right)
 \end{array}
 \end{equation}
 and
\begin{equation}
\label{r32dg}
   r(y)=\frac{16y\left(\sqrt{y}-1\right)^2\left(2\sqrt{3}\,\delta\,y-2\,y+2\sqrt{3}\,\delta\,\sqrt{y}-2\sqrt{y}-\sqrt{3}\,\delta\right)^2}{3\left(\sqrt{3}\,\delta\left({y}^{2}-3y
   +2\sqrt{y}\right)-\left(y-1\right)^2 \right)^2}~.
\end{equation}

Using Eq.~(\ref{equyslr}), we get
\be
\label{dNdy}
\frac{dN}{dy}={\frac {3\,(\,\sqrt{3}\,\delta\,\left(y^2 +2\sqrt {y}- 3y\right)- \left( y-1 \right) ^{2})}{2\,{y}^{3/2} \left(  \left(2 -3\,
\sqrt {3}\,\delta \right) \sqrt{y}+ \sqrt{3}\,\delta\, \left(2\,{y}^{3/2}+1 \right) -2\,{y}^{3/2} \right) }}\,.
\ee

After integration, we obtain the following expression:
\be
\label{Ny}
\begin{array}{cl}
\displaystyle N_e\!&\displaystyle={}N_0-{\frac { \left( 3\,\sqrt {3}\delta-6\,{\delta}^{2}-2 \right)
\ln  \left( y \right) }{2\,{\delta}^{2}}}-{\frac {3\,(\sqrt {3}
\delta-1)}{\delta\,\sqrt {y} \left(\sqrt {3}-3\,\delta \right) }}\\[2.7mm] &\displaystyle
{}+{\frac { 6\,\sqrt {
3}\delta-9\,{\delta}^{2}-4}{4\,{
\delta}^{2}}}\,\ln\left( 2\sqrt {3}\,y-6\,\delta\,y+2\sqrt{3}\sqrt{y}-6\,\delta\sqrt{y}+3\,\delta\right) \\[2.7mm] &\displaystyle
{}-{\frac {3\,\left( 48\,\sqrt {3}{\delta}^{3}-27\,{\delta}^{4}+
18\,\sqrt {3}\delta-81\,{\delta}^{2}-4 \right)}{2\,{\delta}^{2} \left( \sqrt {3}-3
\,\delta \right) \sqrt { \left( \sqrt {3}-3\,\delta
 \right)  \left( -9\,\delta+\sqrt {3} \right) }}}\\[2.7mm] &\displaystyle
\times\mathrm{arctanh}\left( {
\frac { \left( 2\,\sqrt {y}+1 \right)  \left( \sqrt {3}-3\,\delta
 \right) }{\sqrt { \left( \sqrt {3}-3\,\delta \right)  \left(\sqrt {3} -9\,
\delta\right) }}} \right).
\end{array}
\ee
This expression has divergences at $\delta=\sqrt{3}/9$ and $\delta=1/\sqrt{3}$. Therefore, these cases should be considered  separately.

In the case of $\delta=\sqrt{3}/9$, we get the potential
\be
V(y)=\frac{V_0}{3}\left(2y+4\sqrt{y}+3\right)\left(1-\sqrt{y}\right)^2,
\ee
and the inflationary parameters
\begin{equation}
n_s=1-\frac{2\sqrt {y}\left(2\sqrt{y}+1\right) \left(8y^2+12y^{3/2}+12y+19\sqrt{y}+3\right) }{ 3\left(\sqrt{y}-1\right)^{2}\left(2y+4\sqrt{y}+3\right)^2}\,,
\ee
\be
r=\frac {16 y \left( 2\sqrt {y}+1 \right)^4}{3\left( \sqrt {y}-
1 \right) ^{2} \left( 2y+4\sqrt {y}+3 \right) ^{2}}\,.
\end{equation}

The slow-roll evolution equation (\ref{dNdy}) at $\delta=\sqrt{3}/9$ has the following form
\be
\frac{d N_e}{dy}={\frac {3\,(1-y)+2\,(y^2-\sqrt {y}) }{2\,{y}^{3/2} \left( \sqrt {y}-1
 \right)  \left( 2\,\sqrt {y}+1 \right) ^{2}}}
\ee
and gives after integration
\be
\label{Nedeltas3_9}
N_e-N_0={}-{\frac {63}{2}}\,\ln\left( 2\,\sqrt {y}+1 \right) +{\frac {33}{2}}
\,\ln  \left( y \right) +{\frac{27}{2\left( 2\,\sqrt{y}+1\right)}}\, +{\frac{9}{\sqrt{y}}}\,.
\ee

In the case of $\d=1/\sqrt{3}$ that corresponds to $\beta=\sqrt{3}/4$, we obtain the potentials
\begin{equation}
\label{Pot1sqrt3}
V(y)=V_0(1-\sqrt{y})^2,\qquad V(\phi)=V_0\left(1-\mathrm{e}^{{}-\phi/(\sqrt{6}M_{Pl})}\right)^2
\end{equation}
and simple expressions for the slow-roll and inflationary parameters:
\begin{equation}\label{param_slr}
    \epsilon=\frac{y}{3\left(1-\sqrt{y}\right)^2},\qquad \eta=\frac{\sqrt{y}(2\sqrt{y}-1)}{3\left(1-\sqrt{y}\right)^2}
\end{equation}
\begin{equation}\label{param_inf}
 n_s=1-\frac{2\sqrt{y}\left(\sqrt{y}+1\right)}{3\left(\sqrt{y}-1 \right)^{2}} ,\qquad  r=\frac{16y}{3\left(1-\sqrt{y}\right)^2}\,.
\end{equation}

At $\d=1/\sqrt{3}$, the conditions $\epsilon(y_{\mathrm{end}})=1$ and $y_{\mathrm{end}}<1$ gives the following solution:
\begin{equation}
\label{yend}
    y_{\mathrm{end}} = \frac{1}{4}\left(3-\sqrt{3}\right)^2.
\end{equation}

The slow-roll evolution equation (\ref{equyslr}) has the following form
\begin{equation}
y'=\frac{2y^{3/2}}{3\left(\sqrt{y}-1\right)}
\end{equation}
and allows us to express $N_e$ via $y$:
\begin{equation}\label{Neydelta1sqrt3}
    N_e=\frac{3}{\sqrt{y}}+\frac{3}{2}\ln(y)+N_0,
\end{equation}
where the integration constant $N_0$ is fixed by the condition $N_e(y_{\mathrm{end}})=0$:
\begin{equation}\label{N0}
    N_0={}-\frac{3}{\sqrt{y_{\mathrm{end}}}}-\frac{3}{2}\ln(y_{\mathrm{end}}).
\end{equation}

Using Eq.~(\ref{nseta32dg}), we calculate the values of $y_{in}$ for suitable values of $n_s$, namely, we solve equations $n_s(y_{in})=0.961$, $n_s(y_{in})=0.965$, and $n_s(y_{in})=0.969$.
It allows us to calculate the corresponding values of $r(y_{in})$. We calculate the number of e-folding during inflation, using Eq.~(\ref{Neydelta1sqrt3}) for $\delta=1/\sqrt{3}$, Eq.~(\ref{Nedeltas3_9}) for $\delta=\sqrt{3}/9$, and Eq.~(\ref{Ny}) for other values of $\delta$. The result of calculations is presented in Table~\ref{F32inflation}.
One can see that the tensor-to-scalar ratio $r$ is an increasing function of $\d$. The value of $r$ at $\d=1/\sqrt{3}$ is almost four times greater than its value at $\d=0$, that is, in the Starobinsky model.
At the same time, $r<0.036$ for all $0\leqslant\delta\leqslant4\sqrt{3}/9$, so there is no contradiction with the observation data.
The potentials $V(y)$  and ${V}(\phi)$ for the same values of $\delta$ are presented in Fig.~\ref{F32oneparameterVphi}.

\begin{table}[h]
\begin{center}
\caption{Values of $y_{\mathrm{end}}$ and values of $y$, $N_e$ and $r$ corresponding to $n_s=0.961$, $n_s=0.965$, and $n_s=0.969$ for some values of the parameter $\d$.\label{F32inflation}}
\begin{tabular}{|c|c|c|c|c|c|c|}
  \hline
  $\delta$ & $0$ & $0.1$ & $\displaystyle\frac{\sqrt{3}}{9}$ & $\displaystyle\frac{2\sqrt{3}}{9}$ & $\displaystyle\frac{1}{\sqrt{3}}$& $\displaystyle\frac{4\sqrt{3}}{9}$ \\ \hline
  $y_{\mathrm{end}}$ &$0.464$ & $0.460$  & $0.455$ & $ 0.439 $ & $0.402$& $0.299$\\
  ${y_{in,}}_{n_s=0.961}$& $0.0140$& $0.0114$  & $0.00895$ & $0.00479$ & $0.00253$ & $0.00146 $\\
  ${y_{in,}}_{n_s=0.965}$ & $0.0126$ &$0.0101$  & $0.00777$ & $0.00402$ & $0.00209$ &  $0.00120 $\\
  ${y_{in,}}_{n_s=0.969}$ &$0.0112$&$0.00878$  & $0.00660$ & $0.00329$ &$0.00168$&    $0.000968 $ \\
  ${N_{e,}}_{n_s=0.961}$ &$49$& $45$ & $44$ & $45$ & $47$& $48$ \\
  ${N_{e,}}_{n_s=0.965}$ & $55$ & $50$ & $49$ & $51$ & $53$& $54$ \\
  ${N_{e,}}_{n_s=0.969}$ &$62$& $57$  & $56$ & $58$ & $60$& $61$ \\
  $r_{n_s=0.961}$ & $0.0043$& $0.0074$  & $0.010$ & $0.014$ & $0.015$& $0.015$ \\
  $r_{n_s=0.965}$ & $0.0035$     & $0.0061$  & $0.0084$ & $0.0114$ & $0.012$& $0.012$ \\
  $r_{n_s=0.969}$& $0.0027$& $0.0049$  & $0.0068$ & $0.0091$ & $0.0097$& $0.0098 $ \\
  \hline
\end{tabular}
\end{center}
\end{table}

Using values of $y_{in}$ for $\delta=\frac{1}{\sqrt{3}}$, we see that in a good approximation:
\begin{equation}
\sqrt{y}\approx\frac{3}{N_e-N_0}\,.
\label{y(N)}
\end{equation}
The approximation (\ref{y(N)}) allows to present inflationary parameters as follows:
\begin{equation}
\label{nsrN}
  n_s \approx 1-\frac{2}{N_e-N_0}\left[\frac{1+\frac{3}{N_e-N_0}}{\left(1-\frac{3}{N_e-N_0}\right)^2}\right], \quad
  r \approx  \frac{48}{\left(N_e-N_0\right)^2\left[1-\frac{3}{N_e-N_0}\right]^2}\,.
\end{equation}

One can see, that expressions in squared brackets in (\ref{nsrN}) are close to~$1$. Therefore, $r$ is proportional to $(N_e-N_0)^{-2}$ as in the Starobinsky model, but the coefficient of the proportionality is  four times larger.

In the case of $\d=4\sqrt{3}/9$ that corresponds to $\beta=0$, the potentials are
 \begin{equation}
\label{Vy32sd}
     V(y)=\frac{V_0}{3}\left(3+\sqrt{y}\right)\left(1-\sqrt{y}\right)^3\,,
\end{equation}
\begin{equation}
     V(\phi)=\frac{V_0}{3}\left(\mathrm{e}^{\phi/(\sqrt{6}M_{Pl})}-1\right)^3\left(1+3\mathrm{e}^{\phi/(\sqrt{6}M_{Pl})}\right)\mathrm{e}^{-2\sqrt{2/3}\,\phi/M_{Pl}}\,.
\end{equation}
This case has been considered in Ref.~\cite{Ivanov:2021chn} in detail.

\section{Conclusion}

It is well-known~\cite{Barrow:1988xh,Rodrigues-da-Silva:2021jab,Ivanov:2021chn} that adding of the $R^n$ term with $n>2$ to the Starobinsky model results to appearing of a maximum of the corresponding scalar potential at some positive value of the scalar field $\phi$. In this case, inflation demands a fine-tuning of the initial value of $\phi$ that makes such inflationary models unrealistic. In the Starobinsky model, this scalar potential is a monotonically increasing function at positive $\phi$. The generalization with the $R^{3/2}$ term proposed in~\cite{Ivanov:2021chn} has the same property, but is not well-defined for negative values of $R$.

In this paper, the new one-parameter generalization of the Starobinsky model with
\begin{equation}
\label{R32beta}
\begin{array}{ccl}
     F(R)&=&\displaystyle\frac{M^2_{Pl}}{2}\left[\left(1-\frac{3}{8}\delta(4\sqrt{3}-9\delta)\right)R - m^2\d\left(\sqrt{3}-\frac{9}{4}\delta\right)^3    \right. \\&{}+&\displaystyle   \left.
      \frac{\d}{m}\left(R+\left(\sqrt{3}-\frac{9}{4}\delta\right)^2 m^2\right)^{3/2}+ \frac{R^2}{6m^2} \right],
\end{array}
\end{equation}
has been constructed.
For any $0\leqslant\delta\leqslant4\sqrt{3}/9$, this model is consistent with cosmological observations. At $\delta=0$, the proposed model coincides with the Starobinsky inflationary model, at $\delta=4\sqrt{3}/9$, it coincides with the model considered in~Ref.~\cite{Ivanov:2021chn}. Note that the parameters $n_s$ and $r$ do not depend on the scalaron mass $m$, so this parameter is fixed by the observable value of $A_s$ as well as in the  Starobinsky inflationary model. The potential $V(\phi)$ is a monotonically increasing function for all positive values of the scalar field (see Fig.~\ref{F32oneparameterVphi}). So, we do not assume a fine tuning of initial values suitable for inflation.

At $\d=1/\sqrt{3}$, we get the most simple form of the scalar potential (\ref{Pot1sqrt3}) and
\begin{equation*}
    F(R)=\frac{M^2_{Pl}}{2}\left[\frac{5}{8}R   + \frac{\sqrt{3}}{3m}\left[R+\frac{3}{16}m^2\right]^{3/2}+ \frac{R^2}{6m^2}-\frac{3}{64}m^2\right]\,.
\end{equation*}
In this case, the tensor-to-scalar ratio $r$ is almost four times larger than in the Starobinsky model (see Table~\ref{F32inflation} and Eq.~(\ref{nsrN})).

We want to note that the scalar potential (\ref{Vphi1}) corresponding to the model (\ref{R32beta}) is polynomial of the exponential function of $\phi$ for any values of the parameter $\delta$. In the case of the model with the $R^{3/2}$ term, the corresponding potential is more complicated~\cite{Ivanov:2021chn}.

Comparing the model (\ref{R32beta}) with the model with the $R^{3/2}$ term proposed in Ref.~\cite{Ivanov:2021chn}, we want to emphasize that the predictions of inflationary parameters are similar, whereas the post-inflationary evolution of the Universe may differ significantly in these models. The oscillations of $R$ around its zero value characterize post-inflationary evolution in the Starobinsky model (see, for example~\cite{Arbuzova:2011fu,Arbuzova:2021etq}). Similar oscillations of $R$ cannot be reproduced in the model with the $R^{3/2}$ term, but may be possible in the model proposed.
For small values of $R/m^2$, the $F(R)$ function (\ref{R32beta}) has the Taylor series expansion (see Eqs.~(\ref{FRsmall}) and (\ref{Fserias})). The Newtonian limit of such $F(R)$ gravity models and the modeling cluster of galaxies can be considered, using the general approach~\cite{Capozziello:2007ms,Capozziello:2008ny}. For models with the $R^{3/2}$ term, the Newtonian (weak energy) limit and the cosmic evolution have been investigated in Ref.~\cite{Modak:2014yza}.

Modified gravity models can be considered as a gravitational alternative for dark energy~\cite{Nojiri:2010wj}. The Starobinsky model~\cite{Starobinsky:1980te} is not suitable to describe the late-time cosmic acceleration. Different $F(R)$ gravity models have been proposed and studied as an
alternative of dark energy models (see, for example, Refs.~\cite{Martin-Moruno:2008qpc,Modak:2014yza,Starobinsky:2007hu,Appleby:2009uf,Capozziello:2005mj,Borowiec:2006hk,Hu:2007nk,Dev:2008rx,Motohashi:2011wy,delaCruz-Dombriz:2015tye,Odintsov:2017hbk,Oikonomou:2022wuk}). We want specially mention two models: the pure $R^{3/2}$ model~\cite{Martin-Moruno:2008qpc} and the model with the $R^{3/2}$ term~\cite{Modak:2014yza} which can be described by action (\ref{FR32alpha}) with $\beta=0$. In future investigations, we plan to study in detail the post-inflationary evolution of the Universe in the proposed one-parametric extension of the Starobinsky inflationary model.

\ack
The authors are grateful to Sergei~V.~Ketov for useful discussions and are partially supported by the Russian Foundation for Basic Research grant No.~20-02-00411.

\section*{References}

\bibliographystyle{iopart-num}
\bibliography{Bibliography_FR32}{}

\end{document}